\documentclass[final,1p,times]{elsarticle} 
\usepackage{graphicx} 
\usepackage{amssymb} 
\usepackage{amsthm} 
\usepackage{lineno} 

\journal{Nuclear Physics A} 
\begin{document} 

\begin{frontmatter} 


\title{Constraining the onset of viscous hydrodynamics}

\author{Mauricio Martinez$^{a,}${\footnote{Speaker, Quark Matter 2009 (QM09), March 30-April 4, 2009, Knoxville, TN, USA.}} 
and Michael Strickland$^{b}$}

\address{$^a$ Helmholtz Research School, Goethe - Universit\"at Frankfurt am Main, Ruth-Moufang-Str. 1, 60438, Frankfurt, Germany.
\\ $^b$ Physics Department, Gettysburg College, Gettysburg, PA 17325 USA.}

\begin{abstract} 
We derive two general criteria that can be used to constrain the initial time of the onset of 2nd-order conformal viscous hydrodynamics in relativistic heavy-ion collisions. We show this explicitly for 0+1 dimensional viscous hydrodynamics and discuss how to extend the constraint to higher dimensions. 
\end{abstract} 

\end{frontmatter} 

Recent applications of viscous hydrodynamics~\cite{Luzum:2008cw} to bulk physics at RHIC, have shown that estimates of the initial time $\tau_0$ are rather uncertain owing to poor knowledge of the input parameters necessary to perform hydrodynamical simulations. In this work we present two criteria that impose lower bounds on $\tau_0$ by requiring that during all the simulated times, the solutions of viscous hydrodynamics satisfy: (1) positivity of the effective longitudinal pressure $P_L$, and (2) the shear tensor $\Pi$ to be small compared with the isotropic pressure $P$, e.g., $|\Pi|\leqslant P/3$. As a result, the allowed $\tau_0$ is non-trivially related with the initial condition of the shear tensor $\Pi_0$ and the initial energy density $\epsilon_0$. We show this by solving 0+1 dimensional 2nd-order conformal viscous hydrodynamics~\cite{Martinez:2009mf}. Assuming an ideal equation of state, the equations of motion for 0+1 dimensional viscous hydrodynamics are given by: 
\begin{eqnarray}
\label{0+1eqener}
\partial_\tau \epsilon&=&-\frac{4}{3}\frac{\epsilon}{\tau}+\frac{\Pi}{\tau} \, ,\\
\label{0+1eqpi}
\partial_\tau \Pi &=& -\frac{\Pi}{\tau_\pi}
+\frac{4 \eta}{3\, \tau_\pi \tau}-\frac{4}{3\, \tau} \Pi
-\frac{\lambda_1}{2\,\tau_\pi\,\eta^2} \left(\Pi\right)^2 \, ,
\end{eqnarray}
where $\epsilon$ is the energy density, $\Pi$ is the shear tensor component, $\eta$ is the shear viscosity, $\tau_\pi$ is the shear relaxation time, and $\lambda_1$ is a coefficient which arises in complete 2nd-order viscous hydrodynamics~\cite{York:2008rr, Baier:2007ix, Bhattacharyya:2008jc}. The transport coefficients values are estimated from the weak~\cite{York:2008rr,Arnold:2000dr} and strong coupling analysis~\cite{Baier:2007ix,Bhattacharyya:2008jc}. By studying a given set of initial conditions $\{\epsilon_0$, $\Pi_0$, $\tau_0\}$, we can determine if the solution satisfies any of the required criteria. The information about $\epsilon_0$ and $\tau_0$ is encoded in the parameter $k=\tau_0\epsilon_0^{1/4}$ whereas the information for $\Pi_0$ is encoded in $\bar\Pi_0\equiv\Pi_0/\epsilon_0$~\cite{Martinez:2009mf}. 

\begin{figure}[t]
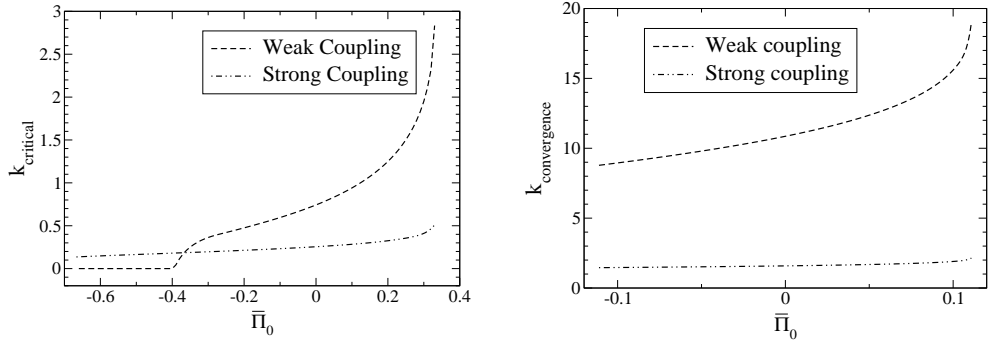

 \hfill
  \begin{minipage}[ht]{.48\textwidth}
  \begin{center}
    \includegraphics[width=6.1cm]{critical.eps}
  \end{center}
  \end{minipage}
  \hfill
  \begin{minipage}[ht]{.48\textwidth}
  \begin{center}
    \includegraphics[width=6.1cm]{convergence.eps}
   \end{center}
  \end{minipage}
\caption{\label{Figure1}Left: Critical boundary in $k$ ($k_{\rm critical}$) as a function of the initial shear, $\overline\Pi_0$. Above the lines, solutions have ${\rm P}_L\geq 0$ at all times.   
Right:  Convergence boundary in $k$ ($k_{\rm convergence}$) as a function of the initial shear, $\overline\Pi_0$. Above these lines, solutions satisfy the convergence constraint.}
\vspace{-0.5cm}
\end{figure}

By requiring positivity of the longitudinal pressure, we find that for a given $\overline\Pi_0$, the system exhibits negative values of $P_L$ if $k$ is below a certain value.  We define this point in $k$ as the ``critical'' value of $k$ or $k_{\rm critical}$.  Above $k_{\rm critical}$, $P_L\geqslant 0$ at all times. The lower bound of $\tau_0$ follows from $\tau_0>\gamma k_{\rm critical} T_0^{-1}$, where $\gamma$ indicates the degrees of freedom of the ideal equation of state and $T_0$ is the initial temperature. In the right panel of Fig.~\ref{Figure1}, we show our critical line in $k_{\rm critical}$ as a funtion of $\bar\Pi_0$ for both coupling regimes. Using this figure, we find that if $T_0$=0.35 GeV and $\bar\Pi_0$=0 are assumed, for strong coupling $\tau_0 > 0.08$ fm/c while in weak coupling $\tau_0 > 0.23$ fm/c~\cite{Martinez:2009mf}.

By imposing on the solution the convergence criteria $|\Pi|\leqslant P/3$, we find that for a given $\overline\Pi_0$, the stronger constraint is not satisfied if $k$ is below a certain value.  We call this point in $k$ as the ``convergence'' value of $k$ or $k_{\rm convergence}$. Above this value of $k=k_{\rm convergence}$, $|\Pi|\leqslant P/3$ is satisfied along all the evolution and therefore, represents a ``reasonable'' simulation. The lower bound follows from $\tau_0>\gamma k_{\rm convergence} T_0^{-1}$. In the left panel of Fig.~\ref{Figure1}, we show our convergence line in $k_{\rm convergence}$ as a funtion of $\bar\Pi_0$ for both coupling limits. From this plot and assuming $T_0$=0.35 GeV and $\bar\Pi_0$=0, we find that for strong coupling $\tau_0 > 0.49$ fm/c whereas for weak coupling $\tau_0 > 3.37$ fm/c~\cite{Martinez:2009mf}. 

For other initial values of $\bar\Pi_0$, one can use Fig.~\ref{Figure1} to establish a lower bound for $\tau_0$ using either the critical~(right panel) or the convergence requirement~(left panel). If one proceeds to more realistic scenarios, e.g. 1+1 and 2+1 viscous hydro, there are new freedoms to consider. The constraints derived here provide guidance for where one might expect 2nd-order viscous hydro to be a good approximation in higher-dimensional cases~\cite{Martinez:2009mf}. 

\vspace*{-0.5cm}
\section*{Acknowledgments}
This work was supported in part by the Helmholtz Research School and the Helmholtz International Center for FAIR Landesoffensive zur Entwicklung Wissenschaftlich-\"Okonomischer Exzellenz program.
\end{document}